\documentclass[aps, prl, twocolumn, showkeys, showpacs, superscriptaddress]{revtex4-1}

\usepackage{amsmath, amssymb, amsthm, dsfont, mathrsfs, bbm}

\usepackage{array}
\usepackage{multirow}	

\usepackage{graphicx}
\usepackage{subfigure}  

\usepackage{url}

\usepackage{color}
	\definecolor{myblue}{rgb}{0,0.3,0.8}
	\definecolor{mygreen}{rgb}{0,0.5,0}
	\definecolor{myblue}{rgb}{0,0.3,0.8}

\usepackage{hyperref} 
	\hypersetup{
		colorlinks, 
		linkcolor = {mygreen}, 
		citecolor = {mygreen}, 
		urlcolor  = {myblue}
	}

\newcommand{\mycommand}[2]{
	\providecommand #1{}
	\renewcommand{#1}{{#2}}
}
\mycommand{\a}{\alpha}
\mycommand{\b}{\beta}
\mycommand{\d}{\delta}
\mycommand{\e}{\varepsilon}
\mycommand{\g}{\gamma}
\mycommand{\G}{\Gamma}
\mycommand{\l}{\lambda}
\mycommand{\r}{\rho}
\mycommand{\s}{\sigma}
\mycommand{\t}{\tau}
\mycommand{\dg}{{\delta g}}
\mycommand{\dt}{{\Delta t}}
\mycommand{\del}{\partial}
\mycommand{\cov}{{\nabla}}
\mycommand{\delvar}{{\overline \partial}}
\mycommand{\sumvar}{{\overline \sum}}
\mycommand{\FF}{{\mathcal F}}
\mycommand{\HH}{{\mathcal H}}
\mycommand{\OO}{{\mathcal O}}
\mycommand{\T}{\textstyle}

\begin{document}

\title{Energy dissipation in flows through curved spaces}

\author{J.-D. Debus} \email{debusj@ethz.ch} \affiliation{ ETH
  Z\"urich, Computational Physics for Engineering Materials, Institute
  for Building Materials, Wolfgang-Pauli-Str. 27, HIT, CH-8093 Z\"urich
  (Switzerland)}
  
\author{M. Mendoza} \email{mmendoza@ethz.ch} \affiliation{ ETH
  Z\"urich, Computational Physics for Engineering Materials, Institute
  for Building Materials, Wolfgang-Pauli-Str. 27, HIT, CH-8093 Z\"urich
  (Switzerland)}

\author{S. Succi} \email{succi@iac.cnr.it} \affiliation{Instituto per le Applicazioni del Calcolo C.N.R., Via dei Taurini, 19 00185, Rome (Italy)}
  
\author{H. J. Herrmann}\email{hjherrmann@ethz.ch} \affiliation{ ETH
  Z\"urich, Computational Physics for Engineering Materials, Institute
  for Building Materials, Wolfgang-Pauli-Str. 27, HIT, CH-8093 Z\"urich
  (Switzerland)}

\begin{abstract}

Fluid dynamics in intrinsically curved geometries is encountered in many physical systems in nature, ranging from microscopic bio-membranes all the way up to general relativity at cosmological scales. Despite the diversity of applications, all of these systems share a common feature: the free motion of particles is affected by inertial forces originating from the curvature of the embedding space. Here we reveal a fundamental process underlying fluid dynamics in curved space: the free motion of fluids, in the complete absence of solid walls or obstacles, exhibits loss of energy due exclusively to the intrinsic curvature of space. We find that local sources of curvature generate viscous stresses as a result of the inertial forces. The curvature-induced viscous forces are shown to cause hitherto unnoticed and yet appreciable energy dissipation, which might play a significant role for a variety of physical systems involving fluid dynamics in curved spaces.



\end{abstract}

\maketitle

\begin{figure}
\centering
\includegraphics[width=\columnwidth]{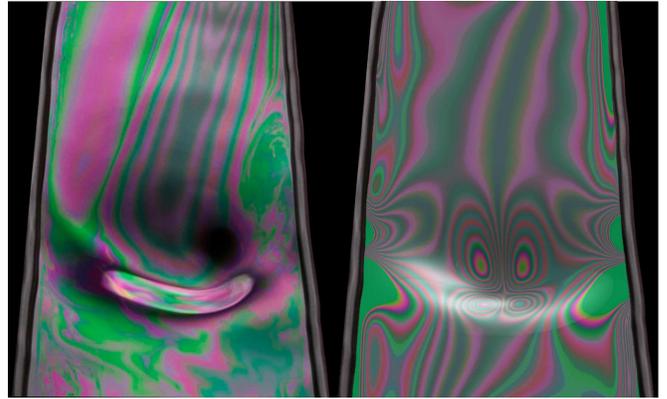}
\caption{\textit{Left:} Experimental realization of flow down a curved soap film. The soap film is spanned between two vertical wires and flows downwards, driven by gravity. The bump in the center is produced by a gentle jet of air (obtained blowing through a straw) inducing a local out-of-plane curvature. 
The colored patterns originate from the thickness variations of the soap film, which are directly correlated with the vorticity field of the flow \cite{rivera1998turbulence}, and thus provide an excellent visualization of the film motion. As can be seen, the incoming laminar stream is considerably distorted at the curved bump. 
\textit{Right:} Numerical simulation of the soap film with a Gaussian curvature bump in the center, showing similar patterns. The colors represent the vorticity field of the flow and correspond to the interference pattern of a real soap film, illuminated by white daylight. As can be seen, there are visible, quadrupol-shaped distortions in the vorticity field next to the curvature source, induced by the inertial forces. Deviations from the experiment originate mainly from the fluid-air interaction which is not taken into account in the simulation. 
}
\label{fig:paper_title_picture}
\end{figure}

In many physical systems, the collective motion of particles can be described by a viscous fluid flow in the hydrodynamic limit. For example, lipid bilayers in microbiology \cite{arroyo2009relaxation,dimova2006practical, cicuta2007diffusion, amar2007stokes}, which constitute the envelope of most of the cell components, behave effectively as a two-dimensional fluid on a curved surface \cite{barrett2015numerical, hu2007continuum, fan2010hydrodynamic,arroyo2009relaxation}. Further applications are curved two-dimensional fluid interfaces, e.g. molecular films around emulsions or aerosol droplets, foam bubbles \cite{scriven1959dynamics, danov2000viscous}, and, on larger scales, the Earth's atmosphere and the photosphere of the sun \cite{priest1982,cao1999navier}. Furthermore, electron transport through curved 2D graphene sheets follows the Navier-Stokes equations in the hydrodynamic regime  \cite{torre2015nonlocal}. Finally, curved soap films provide a daily-life application of 2D flow on curved surfaces \cite{chomaz2001dynamics,zhang2000flexible,vivek2015measuring,reuther2015interplay}. As an illustration, the left half of Fig. \ref{fig:paper_title_picture} shows a soap film, spanned between two vertical wires. With a gentle jet of air we create an out-of-plane curvature, distorting the laminar stream, as can be clearly seen by the colored filaments of the soap film. On the other hand, the right half of Fig. \ref{fig:paper_title_picture} shows our numerical simulation of the soap film experiment, showing similar patterns, where the colors depict the vorticity of the flow, being directly coupled to the thickness of the soap film \cite{rivera1998turbulence}. 
As one can appreciate, the vorticity field is amplified in the center, implying velocity gradients and shear next to the curvature.
In this paper, we analyze this effect in more detail and show that spatial curvature itself generates viscous forces leading to curvature-induced energy dissipation.

\begin{figure}
	\includegraphics[width=0.47\textwidth]{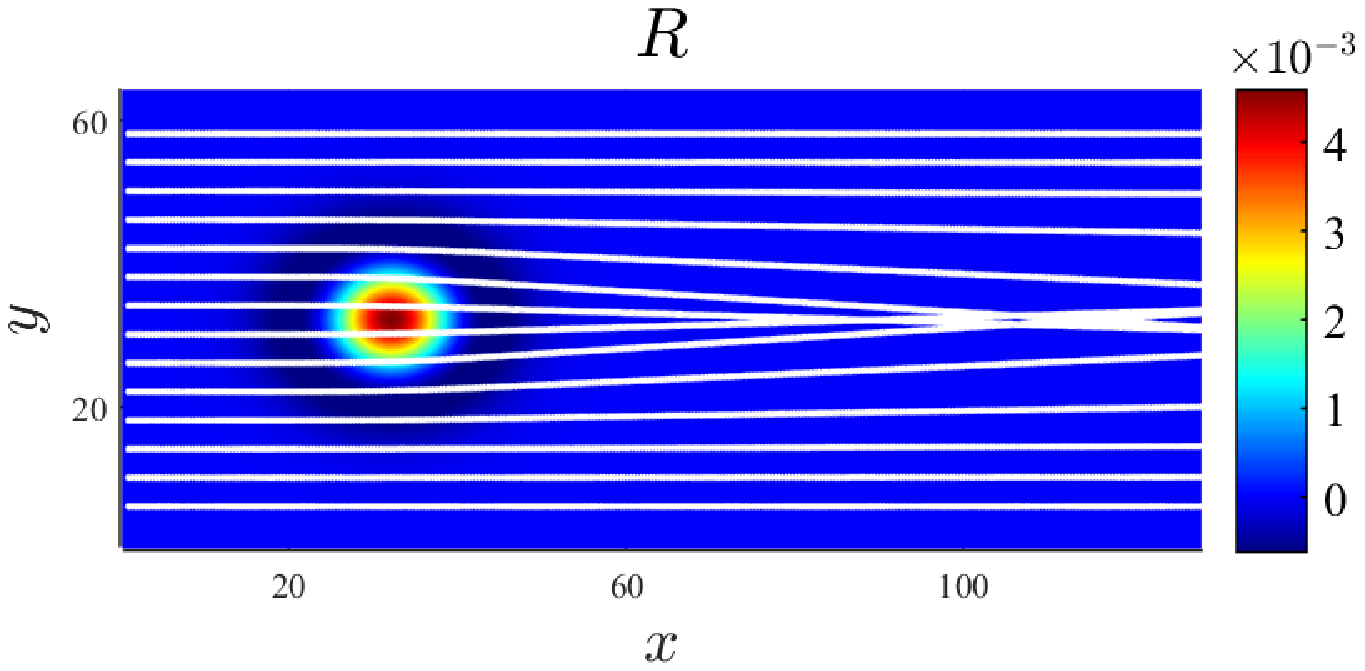}
	~\\
	\includegraphics[width=0.47\textwidth]{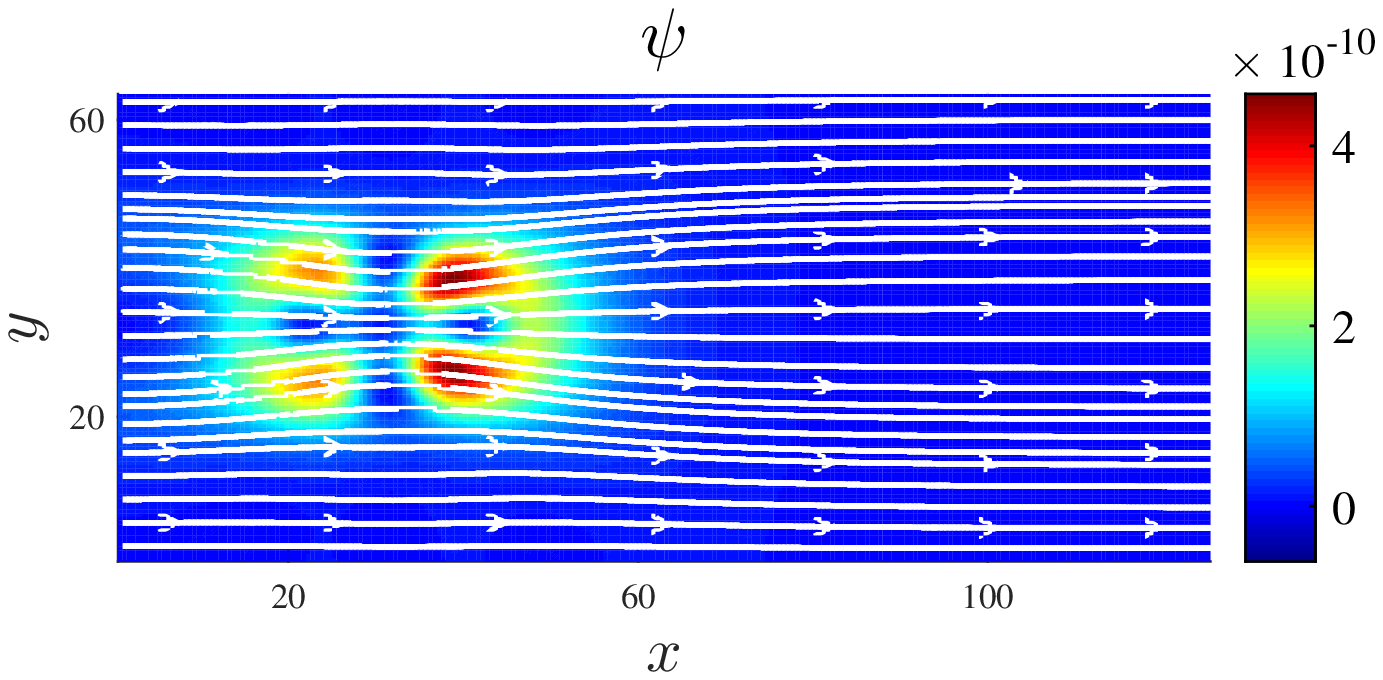}
	\caption{Simulation results for flow past a centered metric perturbation. The metric perturbation is of the shape $\dg = -a_0 \exp (-\|\vec r\|^2/2 r_0^2)$ with amplitude $a_0 = -0.1$ and width $r_0 = 6$, embedded in a rectangular medium of size $L_x \times L_y = 128 \times 64$. The flow is driven by a pressure gradient of $|\nabla P| = 5.8 \times 10^{-9}$, and the kinematic viscosity of the fluid is set to $\nu = 0.185$ (all quantities are given in numerical units). At the inlet ($x=0$) and outlet ($x = L_x$), open boundary conditions have been applied, whereas periodic boundary conditions are used in $y$-direction in order to avoid dissipative effects originating from boundary walls.
\textit{Upper figure:} The Ricci curvature scalar $R$, i.e. the strength of the curvature. The white lines represent the geodesic lines originating parallely from the inlet. As can be seen, the convex curvature field exerts a focusing effect on the geodesics, caused by the attractive inertial force field.
\textit{Lower figure:} The colors represent the local energy dissipation function $\psi = \left( \cov_i u_j \right) \sigma^{ij}$ in the stationary state, where $\sigma^{ij}$ denotes the viscous stress tensor. 
The dissipative effect around the curvature source is clearly visible, and the dissipation function shows a similar quadrupol pattern as the vorticity field in Fig. \ref{fig:paper_title_picture}.
The white lines represent the simulated flow streamlines, which are narrowed around the metric perturbation, thus inducing larger velocity gradients between adjacent fluid layers. The corresponding Reynolds number of the system is defined by $Re = 2 r_0 \Phi / \nu \approx 0.6$, where $\Phi = \frac{1}{S} \int_S \rho\, u^x \sqrt g \, dy$ denotes the mass flow, evaluated on an arbitrary channel cross-section $S$.
} 
\label{fig:paper_single_campylon}
\end{figure}

In order to gain a deeper understanding of the underlying process, we consider a smooth curvature source, described by the metric tensor $g_{ij} = (1 + \dg) \,\delta_{ij}$ with a Gaussian perturbation $\dg(\vec r) = -a_0 \exp (-\| \vec r - \vec r_1 \|^2/2 r_0^2)$, where $\vec r_1$ denotes the center position, $a_0$ the amplitude and $r_0$ the width of the perturbation. The corresponding curvature field can be quantified by the Ricci curvature scalar, as depicted in the upper plot in Fig. \ref{fig:paper_single_campylon}. As one can see, the positive curvature field causes an attractive inertial force field, focusing the geodesic lines $\g(t)$, calculated from the geodesic equation $\cov_{\dot \g} \dot \g = 0$, towards the center.
The simulation of the fluid is based on the covariant Navier-Stokes equations,
\begin{align}
	\nonumber
	&\del_t \rho + \cov_i \left(\rho u^i \right) = 0, \\
 	\label{eq:NS}
	&\del_t \left(\rho u^i \right) + \cov_j \left( \rho u^i u^j \right)
	 = - \cov^i P + \cov_j \s^{ij},
\end{align}
characterizing the time evolution of the fluid density $\rho$ and velocity $u$ on a manifold, equipped with a metric tensor $g_{ij}$ and a covariant derivative $\cov$. Here, $P$ denotes the hydrostatic pressure and $\s^{ij}$ the viscous stress tensor, given by
\begin{align*}
	\sigma^{ij} &= \nu \left( \cov^i (\r\, u^j) + \cov^j (\r\, u^i)  + g^{ij} \cov_k (\r\, u^k)  \right),
\end{align*}
where $\nu$ denotes the kinematic viscosity. We find that the flow, interacting with the curvature source, converges to a stationary equilibrium state after sufficient time, even though we consider an open system where the fluid is continuously driven by the constant pressure drop. The lower plot in Fig. \ref{fig:paper_single_campylon} shows the corresponding velocity streamlines in the stationary state, which are bent towards the center of the metric perturbation by the inertial forces, thus introducing larger velocity gradients $\cov_i u_j$ between adjacent fluid layers. Consequently, viscous stresses are generated within the fluid, explaining the convergence to the steady state, as the viscous forces oppose the pressure drop. At the same time, the presence of viscous stresses causes an irreversible dissipation of energy, which can be measured locally by the dissipation function $\psi = \left( \cov_i u_j \right) \sigma^{ij}$, as depicted in the lower plot in Fig. \ref{fig:paper_single_campylon}. 
The deviation of the velocity streamlines from the geodesic lines is explained by the fact that in a viscous fluid the flow is affected by interparticle collisions, whereas geodesics only represent the trajectories of non-interacting free-falling particles.

In an experiment with a soap film of width $L = 10 \text{cm}$, height $h \approx 1 \mu\text{m}$, mass density $\rho = \approx 10^{-6} \text{g/cm}^2$ and effective 2D dynamic viscosity $\mu \approx 10^{-9} \text{Pa m s}$, nylon threads are used to constrain the soap film, which we account for by applying no-slip boundary conditions, providing an additional source of dissipation \cite{vivek2015measuring,zhang2000flexible}. In order to find the relative magnitude of the curvature-induced dissipation for a metric perturbation of range $r_0 \approx 1 \text{cm}$, we compare the dissipation in a flat soap film to the dissipation in the curved film, finding that the average dissipation increases by about $1\%$ from $\langle \psi_0 \rangle = (6.348 \pm 0.002) \cdot 10^{-8} \text{J/s m}^2$ (flat film) to $\langle \psi \rangle = (6.392 \pm 0.002) \cdot 10^{-8} \text{J/s m}^2$ (curved film). 
This might seem a small correction, but it grows significantly with the number and strength of the metric perturbations, thus leading to non-negligible effects. For a system with 16 randomly arranged metric perturbations, for example, the dissipation increases to about $12 \%$.

\begin{figure}
\centering
\includegraphics[width=\columnwidth]{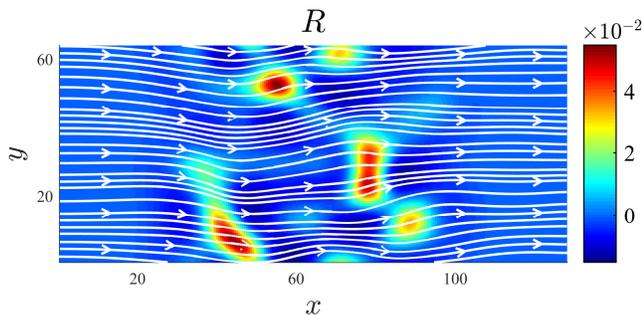}
\caption{Flow past a randomly curved medium. The colors illustrate the Ricci curvature scalar $R$ for an open system with $16$ randomly arranged, Gaussian-shaped metric perturbations of amplitude $a_0 = 0.1$ and width $r_0 = 6$. The white lines represent the streamlines of the flow, driven by a pressure gradient of $|\nabla P| = 5.8 \times 10^{-6}$ with a kinematic viscosity of $\nu = 0.185$ (in numerical units). As can be seen, the flow streamlines are bent by the curvature.}
\label{fig:paper_Ricci2}
\end{figure}

\begin{figure}[ht]
\includegraphics[width=\columnwidth]{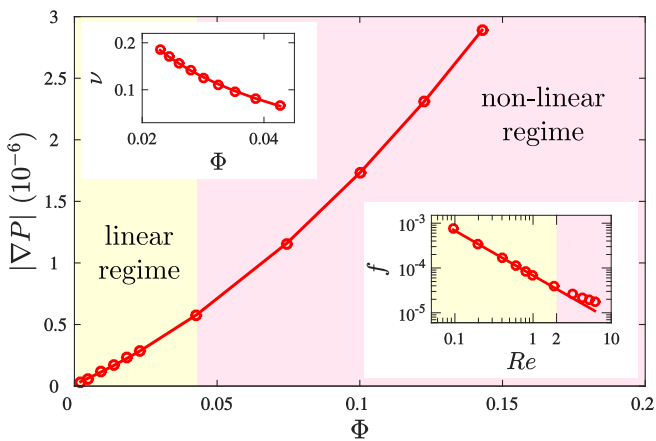}
\caption{Transport law for a curved medium with $16$ metric perturbations (amplitude $a_0=0.1$, width $r_0=4$). The solid line in the main plot represents a fit with respect to the non-linear transport law, $|\nabla P| = \a \nu\, \Phi + \b\, \Phi^2$ for $\a = (5.8 \pm 0.2) \times 10^{-5}$ and $\b = (6.6 \pm 0.3) \times 10^{-5}$. 
\textit{Upper inset:} Flux $\Phi$ as function of viscosity $\nu$ at constant pressure drop $|\nabla P| = 2.9 \times 10^{-7}$. The solid line represents a fit using Eq. (\ref{eq:fit-function}), obtaining $\a = (6.01 \pm 0.06) \times 10^{-5}$ and $\b = (6.7 \pm 0.3) \times 10^{-5}$, in agreement with the fitting values found above. 
\textit{Lower inset:} Friction factor $f = |\nabla P|/\Phi^2 d$ as function of the Reynolds number $Re = \Phi d/\mu$. We observe a linear power law, $f = \a / Re$, for small Reynolds numbers, $Re \lesssim 2$, where the flow is dominated by viscous shear forces, while for higher Reynolds numbers, $Re \gtrsim 2$ the departure from the linear law is visible.}
\label{fig:paper_Darcys_Law_combined}
\end{figure}

\begin{figure}
\includegraphics[width=\columnwidth]{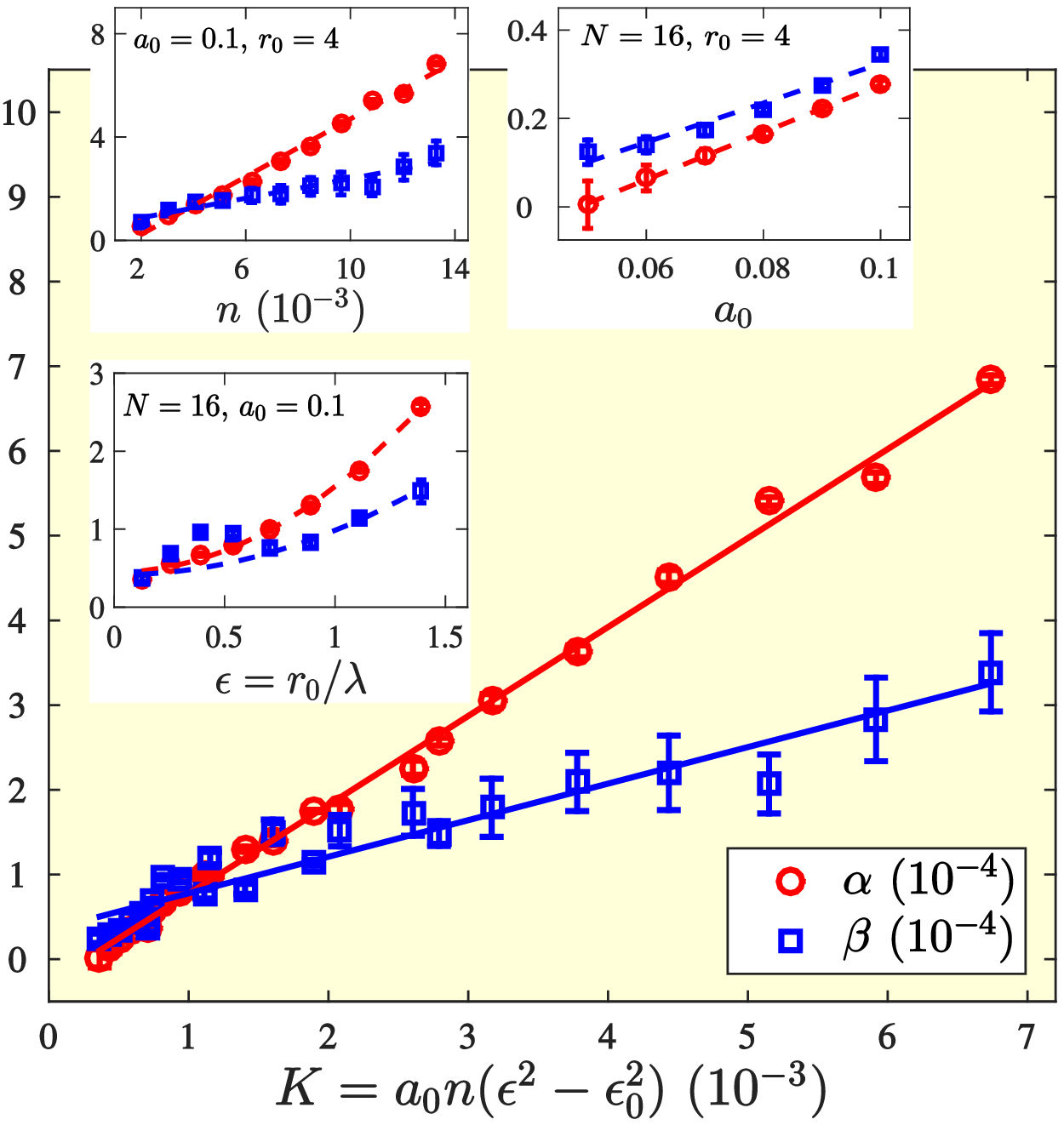}
\caption{Dependence of the reciprocal permeabilities $\a$ and $\b$ on the geometry. \textit{Insets:} $\a$ and $\b$ as function of the amplitude $a_0$, dimensionless range $\e = r_0/\l$ and number density $n = N/V$ of the metric perturbations. \textit{Main plot:} Data collapse when plotting $\a$ and $\b$ as function of the dimensionless average metric perturbation $K = a_0 n (\e^2 - \e_0^2)$ for $\e_0 \approx 1.3$. As can be seen, $\a$ and $\b$ increase linearly with the average metric perturbation, and the solid lines represent linear fits (see appendix for the fitting coefficients). Consequently, the permeability of the medium decreases with increasing strength of the metric perturbation. In all plots, the error bars originate from fitting the non-linear transport law, $|\nabla P| = \a \nu\, \Phi + \b\, \Phi^2$, to our data.}
\label{fig:paper_Darcys_Law_1}
\end{figure}

While in a soap film experiment, the flow is strongly affected by the confining wires, adding a significant source of dissipation to the flow, we are numerically able to study curvature-induced dissipation in open systems without boundaries by using numerical simulations. Fig. \ref{fig:paper_Ricci2} depicts the flow through an open system with 16 randomly arranged metric perturbations, where the velocity streamlines are bent only by the intrinsic curvature. 
A striking observation is made by plotting the pressure gradient, $|\nabla P|$, as function of the total mass flow $\Phi$, as depicted in Fig. \ref{fig:paper_Darcys_Law_combined}. As can be seen, for $\Phi \lesssim 0.04$, there is a linear correlation between $|\nabla P|$ and $\Phi$, where the flow is governed by the terms linear in the velocity in the Navier-Stokes equation (\ref{eq:NS}). These terms represent curvature-induced viscous forces, $\nabla_j \s^{ij} \approx \nu \cov_j \cov^j (\r u^i)$, originating from the viscous stress tensor and thus from the shear $\cov_j u^i$ induced between adjacent fluid layers. For stronger mass flows, $\Phi \gtrsim 0.04$, the transport law in Fig. \ref{fig:paper_Darcys_Law_combined} requires a non-linear correction, originating from the terms quadratic in the velocity $\sim \nabla_j (\r u^i u^j)$. In this regime, the quadratic terms $\sim \G^i_{jk} \r u^j u^k$ begin to dominate over the viscous forces, causing a break down of the linear law. This becomes clearer when plotting the ratio between the shear stress and the kinetic energy, expressed by the friction fractor $f = |\nabla P|/\Phi^2 d$, as function of the Reynolds number $Re = 2 r_0 \Phi / \nu$, as depicted in the lower inset of Fig. \ref{fig:paper_Darcys_Law_combined}. 
Finally, the upper inset of Fig. \ref{fig:paper_Darcys_Law_combined} depicts the  dependence of the flux on the viscosity, measured at constant pressure drop, finding an anti-correlation between $\Phi$ and $\nu$. 
All these results can be summarized in the following transport law, describing both the linear regime and the non-linear effects:
\begin{align}\label{eq:fit-function}
	|\nabla P| = \a \nu\, \Phi + \b\, \Phi^2,
\end{align}
where $\a$ and $\b$ are parameters, which can be interpreted as reciprocal permeabilities, and $\nu$ denotes the kinematic viscosity. 
Interestingly, Eq. (\ref{eq:fit-function}) is similar to the Darcy-Forchheimer's law for porous media \cite{darcy1856,dullien2012porous}, where the dissipation originates from the interaction of the flow with solid obstacles. However, there are major physical differences regarding the underlying mechanisms in the two types of media. In a porous medium, the obstacles are impermeable solid structures, while the fluid motion is governed by the Navier-Stokes equation in flat space. In a curved medium, however, the background space itself is curved, inducing inertial forces, viscous stresses and thus dissipation. Furthermore, since the metric perturbations are permeable to the fluid, the flow typically does not form sharp channel-like paths as in a porous medium.

In order to formulate a transport law for general geometries, we study the dependence of the coefficients $\a$ and $\b$ on the parameters of the curved space. We have performed simulations for a wide range of metric parameters by varying the amplitude $a_0$, the range $r_0$ and the number density $n = N/V$ of the metric perturbations, where $V$ denotes the volume of the curved space. We observe that all the data collapse onto a single common curve when plotting $\a$ and $\b$ as function of the non-dimensionalized average metric perturbation $K = a_0 n (\e^2 - \e_0^2)$, where $\e = r_0/\l$ denotes the non-dimensionalized perturbation range, normalized by the characteristic distance between the perturbations, $\lambda = \sqrt{V/N}$, and $\e_0 \approx 1.3$. The resulting curves are depicted in Fig. \ref{fig:paper_Darcys_Law_1}, showing a linear increase of $\a$ and $\b$ with $K$. Accordingly, the permeabilities $\sim \a^{-1}, \b^{-1}$ decrease with the average strength of the metric perturbation. This behavior agrees with our expectation, since the stronger the metric perturbations in the medium, the larger the effective resistance against the flow will be.

Summarizing, we have presented a fundamental physical process inherent to fluid dynamics in curved space, namely curvature-induced dissipation. 
We have shown that local sources of curvature generate viscous stresses, which tend to narrow or widen the streamlines and thus introduce velocity gradients between adjacent fluid layers, leading to an irreversible dissipation of energy.
In order to demonstrate the far-reaching consequences of curvature-induced dissipation, we have studied media with randomly-distributed curvature sources. We have observed that the flow converges to a stationary equilibrium state solely due to curvature-induced dissipation.
As a consequence, the flow has been found to satisfy 
a non-linear transport law, 
$|\nabla P| = \a \nu\, \Phi + \b\, \Phi^2$, where the reciprocal permeabilities $\a$ and $\b$ are linearly correlated with the average metric perturbation in the medium.

The present work might have important implications on physical systems described by hydrodynamics on curved manifolds. For example, curvature-induced dissipation is expected to influence the two-dimensional fluidity and thus the biological function of curved lipid bilayers, constituting the cell membranes. In particular, the diffusive motion of the tracer particles used to measure the membrane viscosity \cite{hormel2014measuring} is probably affected by the curvature.
Furthermore, curvature-induced dissipation might influence astrophysical hydrodynamic processes, such as gas dynamics in stars or clouds, and might even play a role in controversial cosmological questions concerning e.g. the fluid dynamical description of dark matter and dark energy \cite{gibson2000turbulent,bento2002generalized} or the energy conservation in the universe.

\section{Acknowledgements}
  We acknowledge financial support from the European Research Council (ERC) Advanced
  Grant 319968-FlowCCS. We further thank Troy Shinbrot, Jun Zhang, Eric Weeks and Damian Berger for useful help and discussions concerning the experimental realizations of soap films.  
  
\section{Author contributions}
All authors have contributed equally in designing the experiment, analyzing the results, and writing the manuscript. J.-D.D. has performed the simulations and the experiment of the soap film.

\section{Appendix}

\subsection{Method}\label{sec:CE}

\subsubsection{Lattice Boltzmann method in curved space}

Flow through curved space is modeled by the lattice Boltzmann (LB) method on manifolds \cite{mendoza2013,mendoza2014,debus2016poiseuille}, based on the LB equation
\begin{align}\label{eq:CE-LB}
	f_\l(x + c_\l \dt, t + \dt) - f_\l(x, t) 
	 = \mathcal C_\l(x,t) + \dt \,F_\l(x,t).
\end{align}
This equation describes the evolution of the distribution function $f_\l$ in space and time according to kinetic theory on manifolds, characterized by the metric tensor $g_{ij}$. All quantities are discretized on the two-dimensional $D2Q17$-lattice, consisting of $17$ lattice vectors $\{c_\l\}_{\l=1}^{17}$ per node.
While the left-hand side of the LB equation represents free streaming, the right-hand side accounts for collisions and inertial forces on the manifold. 
Collisions are described by the BGK collision operator $\mathcal C_\l = - ( f_\l - f_\l^{\rm eq} )/\t$ \cite{BGK}, where $\t$ is the relaxation time and $f_\l^{\rm eq}$ the Maxwell-Boltzmann equilibrium distribution in curved space. On the other hand, the inertial forces on the manifold enter the LB equation through the forcing term $F_\l$.
In the hydrodynamic limit, the LB equation flows into the covariant Navier-Stokes conservation equations,
\begin{align*}
	\del_t \rho + \cov_i \left(\rho u^i \right) = 0, \qquad
	\del_t \left(\rho u^i \right) + \cov_j T^{ij} = 0,
\end{align*}
where the macroscopic fluid density $\rho$ and fluid velocity $\vec u$ are obtained by taking the moments of the distribution function,
\begin{align*}
	\sumvar_\l f_\l = \rho
	\quad,\quad
	\sumvar_\l c_\l^i f_\l = \rho u^i.
\end{align*}
Here, the generalized sum $\sumvar_\l := \sqrt g\, \sum_\l$ represents the (discretized) momentum-space integral in curved space, where the factor $\sqrt g$, the square root of the metric determinant, originates from the volume element on general manifolds. 
The energy-stress tensor $T^{ij} = \Pi^{{\rm eq},ij} - \sigma^{ij}$ is composed of the free momenum-flux tensor $\Pi^{{\rm eq},ij}$ and the viscous stress tensor $\s^{ij}$:
\begin{align*}
\Pi^{{\rm eq},ij} &= \sumvar_\l c_\l^i c_\l^j f_\l^{\rm eq} = P g^{ij} + \rho u^i u^j \\
\sigma^{ij} &= - \T\left( 1 - \frac{1}{2 \t}\right) \sumvar_\l  c_\l^i c_\l^j \left( f_\l - f_\l^{\rm eq} \right) \\
	&= \nu \left( \cov^i (\r u^j) + \cov^j (\r u^i)  + g^{ij} \cov_k (\r u^k)  \right),
\end{align*}
where $P = \rho\, \theta$ denotes the hydrostatic pressure, $\theta$ the normalized temperature (we work in the isothermal limit $\theta = 1$), $g^{ij}$ the inverse metric tensor, $\nu = c_s^2 (\t - \frac{1}{2})\, \dt$ the kinematic viscosity, and $c_s = \sqrt{\frac{25 - \sqrt{193}}{30}}$ the lattice specific speed of sound.

The equilibrium distribution $f_\l^{\rm{eq}}$ as well as the forcing term $F_\l$ are expanded into an orthonormal polynomial basis (tensor Hermite polynomials \cite{grad1949}) in order to recover their macroscopic moments exactly up to third order through Gauss-Hermite quadrature.
Explicitly, the tensor Hermite polynomials are given by
\begin{align*}
	\HH_{(0)}(v) &= 1, \quad
	\HH_{(1)}^i(v) = v^i, \quad
	\HH_{(2)}^{ij}(v) = v^i v^j - \d^{ij}, \\
	\HH_{(3)}^{ijk}(v) &= v^i v^j v^k - 
	\left( \delta^{ij} v^k 
	+ \delta^{jk} v^i 
	+ \delta^{ki} v^j \right)
\end{align*}
and satisfy the orthogonality relation up to third order (for the given third-order $D2Q17$-lattice):
\begin{align*}
	\sum_\l w_\l\ \HH_{(n),\l}^{I_n} \cdot \HH_{(m),\l}^{J_m} = \d_{nm} \d^{I_n J_n},
\end{align*}
where $n,m = 0,1,2,3$ and $\HH_{(n),\l}^{I_n} := \HH_{(n)}^{I_n}\left( c_\l / c_s\right)$. Here, $w_\l$ and $c_\l$ are the lattice weights and  velocities, respectively, (listed in Table \ref{tab:d2q17-lattice}), $c_s = \sqrt{\frac{25 - \sqrt{193}}{30}}$ denotes the lattice specific speed of sound, $I_n = (i_1,\ldots,i_n)$ and $J_m = (j_1,\ldots,j_m)$ are index tuple and $\d^{I_n J_n} := \sum_{\s \in S_n}(\d^{i_1 j_{\s(1)}} \cdots \d^{i_n j_{\s(n)}})$ denotes the fully symmetric tensorial Kronecker delta.

\begin{table}[ht]
  \centering 
  \begin{tabular}{|@{\quad}c@{\quad}|@{\quad}c@{\quad}|@{\quad}c@{\quad}|}\hline
    $\l$ & $c_\l$ & $w_\l$ \\ \hline\hline
    1 & $(0,0)$ & $\frac{575+193 \sqrt{193}}{8100}$ \\[2pt] \hline
    2-3 & $(\pm 1,0)$ & \multirow{2}{*}{$\frac{3355-91 \sqrt{193}}{18000}$} \\
    4-5 & $(0,\pm 1)$ &  \\ [2pt] \hline
    6-9 & $(\pm 1,\pm 1)$ & $\frac{655+17 \sqrt{193}}{27000}$ \\ [2pt] \hline
    10-13 & $(\pm 2,\pm 2)$ & $\frac{685-49 \sqrt{193}}{54000}$ \\ [2pt] \hline
    14-15 & $(\pm 3,0)$ & \multirow{2}{*}{$\frac{1445-101 \sqrt{193}}{162000}$} \\ 
    16-17 & $(0,\pm 3)$ & \\ \hline
  \end{tabular}
  \caption{Discrete velocity vectors $c_\l$ of the D2Q17 lattice and the
  corresponding weights $w_\l$ in the Hermite expansion.} 
  \label{tab:d2q17-lattice}  
\end{table}

The expanded equilibrium distribution reads
\begin{align}
	\label{eq:CE-feq-expansion}
	f_\l^{\rm eq}(x,t) &= \frac{w_\l}{\sqrt g} \sum_{n=0}^{3} \frac{1}{n!\,c_s^n}\, a_{(n)}^{{\rm eq},I_n}(x,t)\, \HH_{(n),\l}^{I_n},	
\end{align}
where the expansion coefficients are given by
\begin{align*}
	&a_{(0)}^{{\rm eq}} = \r,\qquad
	a_{(1)}^{{\rm eq},i} = \r u^i,\qquad 
	a_{(2)}^{{\rm eq},ij} = \r c_s^2 \Delta^{ij} + \r u^i u^j, \\
	&a_{(3)}^{{\rm eq},ijk} = \r c_s^2 \left( \Delta^{ij} u^k + \Delta^{jk} u^i +
	\Delta^{ki} u^j \right) + \r u^i u^j u^k,
\end{align*}
where $\Delta^{ij} := g^{ij} - \delta^{ij}$.
The corresponding moments read
\begin{align}
	\label{eq:CE-eq-moments1}
	\r &= \sumvar_\l f^{\rm eq}_\l, \\	
	\label{eq:CE-eq-moments2}
	\r u^i &= \sumvar_\l f^{\rm eq}_\l c_\l^i, \\
	\label{eq:CE-eq-moments3}
	\Pi^{{\rm eq},ij} &= \sumvar_\l f^{\rm eq}_\l c_\l^i c_\l^j = \r\left( c_s^2 g^{ij} + u^i u^j \right),\\
	\nonumber
	\Sigma^{{\rm eq},ijk} &= \sumvar_\l f^{\rm eq}_\l c_\l^i c_\l^j c_\l^k = \r c_s^2 \left( u^i g^{jk} + u^j g^{ik} + u^k g^{ij}  \right)\\
	\label{eq:CE-Sigma}
	&\qquad\qquad\qquad\qquad\quad+ \r u^i u^j u^k.
\end{align}

The inertial forces on the manifold enter the LB equation through the forcing term $F_\l$, which needs special treatment in order to cancel spurious discrete lattice effects in the hydrodynamic equations (cf. \cite{Guo2002}). To this end, we employ the trapezoidal rule for the time integration in the LB equation by using an improved forcing term:
\begin{align*}
	F_\l(x, t) := \FF_\l(x, t)
	+ \T\frac{1}{2} \left( \FF_\l(x + c_\l \dt, t ) 
	- \FF_\l(x, t - \dt) \right),
\end{align*}
where $\FF_\l$ is expanded in Hermite polynomials,
\begin{align}
	\label{eq:CE-F-expansion}
	\mathcal F_\l(x,t) 
	&= \frac{w_\l}{\sqrt g} 
	\sum_{n=0}^{2} \frac{1}{n!\,c_s^n}\, b_{(n)}^{I_n}(x,t)\, \HH_{(n),\l}^{I_n},	
\end{align}
with expansion coefficients
\begin{align*}
	b_{(0)} &= 
	- \G^i_{ij} \r u^j 
	- \G^j_{ij} \r u^i,\\
	b_{(1)}^{i} &= 
	- \G^i_{jk} T^{jk} 
	- \G^j_{jk} T^{ik} 
	- \G^k_{jk} T^{ji},\\
	b_{(2)}^{ij} &= 
	-\G^i_{kl} \Sigma^{{\rm eq},jkl}
	-\G^j_{kl} \Sigma^{{\rm eq},ikl}
	-\G^k_{kl} \Sigma^{{\rm eq},ijl}
	-\G^l_{kl} \Sigma^{{\rm eq},ijk}\\
	&\ \ \ - c_s^2 \delta^{ij} b_{(0)}.
\end{align*}
The corresponding moments of the forcing term are given by
\begin{align}
	\label{eq:CE-A}
	A &= \sumvar_\l \FF_\l = 
	- \G^i_{ij} \r u^j 
	- \G^j_{ij} \r u^i, \\
	\label{eq:CE-B}
	B^i &= \sumvar_\l \FF_\l c_\l^i = 
	- \G^i_{jk} T^{jk} 
	- \G^j_{jk} T^{ik} 
	- \G^k_{jk} T^{ji},\\
	\nonumber
C^{ij} &= \sumvar_\l \FF_\l c_\l^i c_\l^j = 
	-\G^i_{kl} \Sigma^{{\rm eq},jkl}
	-\G^j_{kl} \Sigma^{{\rm eq},ikl}\\
	\label{eq:CE-C}
	&\quad\qquad\qquad\qquad\ \ 
	 -\G^k_{kl} \Sigma^{{\rm eq},ijl}
	-\G^l_{kl} \Sigma^{{\rm eq},ijk}.
\end{align}

With the improved forcing term, the Navier-Stokes equations are recovered at second order in space and time, as will be shown in the following section. This significantly improves the accuracy of the results.
The advantage of this force correction scheme, as compared to the commonly used scheme by Guo \textit{et al.} \cite{Guo2002}, is that the macroscopic moments (density $\rho$, velocity $u$ etc.) are not modified and thus retain their 
original physical meaning. Moreover, for the special forcing term used in our model, a modification of the macroscopic moments according to Ref. \cite{Guo2002} would require solving a complicated system of equations for $\rho$ and $u$. 
With our method, such modifications of the macroscopic moments are dispensed by employing the trapezoidal rule for the time integration in the lattice Boltzmann equation.

\subsubsection{Simulation setups}

For the simulations performed for this Letter, we use a discretization step $\dt = 0.5$ (corresponding to $256 \times 128$ grid points) and set the fluid viscosity to $\nu = c_s^2/2 \approx 0.185$, which fixes the relaxation parameter $\t$ accordingly. 
At the inlet and outlet, we use a non-equilibrium extrapolation method, Ref. \cite{zhao2002non}, to impose open boundaries with a pressure gradient, driving the fluid towards the outlet. No-slip boundaries (walls) are implemented by imposing zero flow velocity in the equilibrium distribution.

For the soap film simulation depicted in Fig. 1 in the Letter, a different set of simulation parameters is used: The channel is modeled by a grid of $512 \times 128$ lattice points with an out-of-plane Gaussian curvature of amplitude $a_0=15$ and width $r_0=28$ (all quantities are given in numerical units), using open boundaries at the top and bottom of the channel, and no-slip boundary conditions at the wires. The flow is driven by a pressure gradient of $\nabla P = 9.2 \times 10^{-10}$, and the vorticity field is measured with a finite difference method at time $t = 1800$. The colorbar is taken from Ref. \cite{colorpy} and corresponds to the real thickness-dependent colors of a soap film with refraction index $n=1.33$, illuminated by white daylight (Illuminant D65).

\subsubsection{Chapman-Enskog expansion}\label{sec:CE}

In the following section, we prove that in the hydrodynamic limit, our curved-space lattice Boltzmann equation flows into the covariant Navier-Stokes equations with second order accuracy in space and time. 
To this end, we will use the following abbreviations for readability:
\begin{align*}
	\sumvar_\l := \sqrt g \, {\sum}_\l 	
	, \qquad
	D_t := \del_t + c_\l^i \del_i
	, \qquad
	\delvar_i := \del_i - \G^j_{ij},
\end{align*}
where $\sqrt g$ denotes the square root of the metric determinant, $\G^i_{jk} = \frac{1}{2} g^{im} ( \del_j g_{km} + \del_k g_{jm} - \del_m g_{jk} )$ the Christoffel symbols, and $c_\l^i$ the lattice vectors.

We perform a Chapman-Enskog multiscale expansion by expanding the distribution function as well as the time and space derivatives in terms of the Knudsen number $\e$ \cite{wolf2000lattice,buick2000gravity}:
\begin{gather*}
	f = f^{(0)} + \e f^{(1)} + \e^2 f^{(2)} + ... , \\
	\del_t = \e \del_t^{(1)} + \e^2 \del_t^{(2)} + ... , \\
	(\del_i,\FF, A, B^i, C^{ij}) = \e (\del_i^{(1)}, \FF^{(1)}, A^{(1)}, B^{(1),i}, C^{(1),ij})
\end{gather*}
Plugging everything into Eq. (\ref{eq:CE-LB}) and comparing orders of $\e$, we obtain the following equations:
\begin{align}
	\label{eq:CE-0}
	&\T\OO(\e^0): \quad
	f_\l^{(0)} = f_\l^{{\rm eq}}, \\ 
	\label{eq:CE-I}
	&\T\OO(\e^1): \quad
	\T D_t^{(1)} f_\l^{(0)} = - \frac{1}{\t \dt} f_\l^{(1)} + \FF_\l^{(1)},\\	
	\label{eq:CE-II}
	&\T\OO(\e^2): \quad
	\T\del_t^{(2)} f_\l^{(0)} + \left(1- \frac{1}{2\t}\right) \,D_t^{(1)} f_\l^{(1)}
	\T= - \frac{1}{\t \dt} f_\l^{(2)}.
\end{align}

\medskip
\textbf{Moments of Eq. (\ref{eq:CE-I}-\ref{eq:CE-II})}

Taking the moments of Eq. (\ref{eq:CE-I}) yields:
\begin{align}
	\label{eq:CE-1}
	\T\sumvar_\l (\ref{eq:CE-I})  :
	\T\quad &\del_t^{(1)} \r + \delvar_i^{(1)} \left(\r u^i \right)
	\T= A^{(1)}, \\
	\label{eq:CE-2}
	\T\sumvar_\l c_\l^i\, (\ref{eq:CE-I})  : \quad &
	\T\del_t^{(1)} \left( \r u^i \right) + \delvar_j^{(1)} \Pi^{(0),ij}
	\T= B^{(1),i},
\end{align}
where $\Pi^{(0),ij} = \Pi^{{\rm eq},ij} = \r\left( c_s^2 g^{ij} + u^i u^j \right)$, and the $\delvar_i$ derivative originates from
\begin{align*}
	{\sumvar}_\l c_\l^i \del_i f_\l^{(0)}
	&= \del_i {\sum}_\l \sqrt g\, c_\l^i f_\l^{(0)}
	-{\sum}_\l  (\del_i \sqrt g)\, c_\l^i f_\l^{(0)}\\
	&= \del_i (\r u^i) - \G^j_{ij} (\r u^i) =: \delvar_i (\r u^i),
\end{align*}
where we have used the identity $\del_i \sqrt g = \G^j_{ij} \sqrt g$.
The moments of Eq. (\ref{eq:CE-II}) are given by
\begin{align}
	\label{eq:CE-3}	
	\T\sumvar_\l (\ref{eq:CE-II}) : \quad&
	\T\del_t^{(2)} \r = 0, \\	
	\label{eq:CE-4}
	\T\sumvar_\l c_\l^i\, (\ref{eq:CE-II}) : \quad&
	\T\del_t^{(2)} \left(\r u^i \right) 
	\T= \delvar_j^{(1)} \s^{(1),ij},
\end{align}
where $\s^{(1),ij}$, the viscous stress tensor (rescaled by $\e$), is defined as
\begin{align*}
	\T\s^{(1),ij} = - \left(1 - \frac{1}{2\t}\right) \sumvar_\l c_\l^i c_\l^j f^{(1)}_\l .
\end{align*}
The explicit expression for the viscous stress tensor will be derived later.

\medskip
\textbf{Continuity Equation}

For the continuity equation, we add $\e \cdot (\ref{eq:CE-1})$ and $\e^2 \cdot (\ref{eq:CE-3})$:
\begin{align*}
	\T\quad\del_t \r +  \delvar_i \left(\r u^i \right)
	= A
\end{align*}
After inserting the explicit expression for $A$ (\ref{eq:CE-A}), we obtain the correct continuity equation:
\begin{align*}
	\del_t \r + \cov_i \left(\r u^i \right) = 0,
\end{align*}
where $\cov$ denotes the covariant derivative.

\medskip
\textbf{Momentum Equation}

Adding $\e \cdot(\ref{eq:CE-2})$ and $\e^2 \cdot (\ref{eq:CE-4})$ yields the momentum conservation equation:
\begin{align*}
	\T \del_t \left( \r u^i \right) 
	+ \delvar_j \Pi^{(0),ij}
	\T= \delvar_j \s^{ij} + B^i.
\end{align*}
Inserting the explicit expression for $B^i$ (\ref{eq:CE-B}) and $\Pi^{(0),ij} = \Pi^{{\rm eq},ij}$ (\ref{eq:CE-eq-moments3}) yields the familiar Navier-Stokes equation:
 \begin{align*}
	\del_t \left( \r\, u^i \right) + \cov_j \left( \r\, u^i u^j + \r \,c_s^2 g^{ij} \right) 
	&= \cov_j \s^{ij}.
\end{align*}

\medskip
\textbf{Viscous Stress Tensor}

For the derivation of the viscous stress tensor $\s^{ij}$, we rewrite
\begin{align*}
	\s^{ij} =\ \ &\T -\left( 1 - \frac{1}{2\t}\right) 
	\e \sumvar_\l c_\l^i c_\l^j f_\l^{(1)} \\
	\overset{(\ref{eq:CE-I})}=& \T \left( \t - \frac{1}{2}\right) \dt\, \e\, \sumvar_\l c_\l^i c_\l^j  \left( D_t^{(1)} f_\l^{(0)} - \FF_\l^{(1)} \right) \\
	\approx\ \ & \T \left( \t - \frac{1}{2}\right) \dt \left( \delvar_k \Sigma^{{\rm eq},ijk} - C^{ij} \right),
\end{align*}
where we have assumed that $\left( \t - \frac{1}{2} \right) \dt\, \del_t^{(1)} \Pi^{(0),ij} \ll 1$.
After plugging in the explicit expressions for $\Sigma^{{\rm eq},ijk}$ (\ref{eq:CE-Sigma}) and $C^{ij}$ (\ref{eq:CE-C}), we obtain the familiar expression for the viscous stress tensor:
\begin{align*}
	\s^{ij} &=  \T \nu \left( \cov^j (\r u^i) + \cov^i (\r u^j)  + g^{ij} \cov_k (\r u^k)  \right),
\end{align*}
where we have defined $\nu := \left( \t - \frac{1}{2}\right) \dt\, c_s^2$ and neglected terms of the order $\OO(u^3) \sim \OO(\text{Ma}^3)$, $\text{Ma}$ being the Mach number.

\subsubsection{Finite resolution study}

\begin{figure}[ht]
\includegraphics[width=\columnwidth]{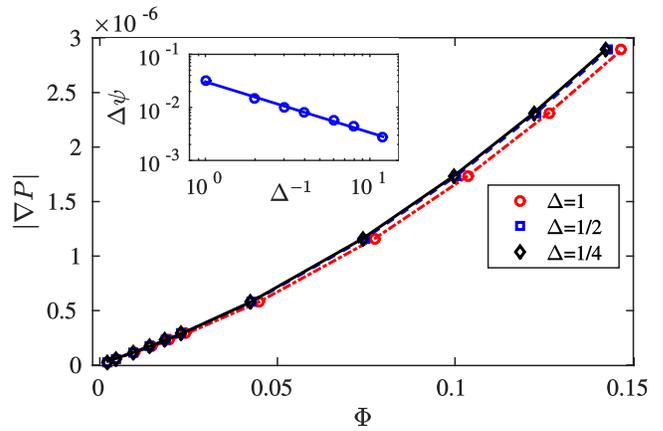}
\caption{Transport law for a curved medium with $16$ metric perturbations ($a_0=0.1$, $r_0=4$) for different discretization steps $\Delta$. \textit{Inset:} Relative error of the total dissipation function, $\Delta \psi$, as function of the grid resolution $\Delta^{-1}$ for a single metric perturbation ($a_0 = -0.1$, $r_0 = 6$).}
\label{fig:paper_Darcys_Law_finite_resolution_combined}
\end{figure}

In order to proof the physicality of our lattice Boltzmann simulations, we have performed a finite resolution study. To this end, we have measured the transport law, $|\nabla P| = \a \nu\, \Phi + \b\, \Phi^2$, for a curved medium with $16$ metric perturbations ($a_0=0.1$, $r_0=4$) for different grid resolutions $\Delta^{-1}$, as depicted in Fig. \ref{fig:paper_Darcys_Law_finite_resolution_combined}. 
As can be seen, the distance between the curves decreases rapidly with increasing resolution, which can be quantified by the relative $L^2$-difference, being less than $1 \%$ between the dashed ($\Delta = \frac{1}{2}$) and solid ($\Delta = \frac{1}{4}$) curve.
Furthermore, we measured the finite resolution effects on the average dissipation, $\langle \psi \rangle = \frac{1}{V} \int \psi\, dV$, measured at time $t = 10^5$ for a single metric perturbation of amplitude $a_0 = -0.1$ and width $r_0 = 6$. In the inset of Fig. \ref{fig:paper_Darcys_Law_finite_resolution_combined}, the relative error 
$\Delta \psi = | \langle \psi \rangle - \langle \psi \rangle_\infty | / \langle \psi \rangle_\infty$ is depicted as function of the resolution $\Delta^{-1}$, where $\langle \psi \rangle_\infty = (2.991 \pm 0.002) \times 10^{-9}$ has been determined by extrapolation. As can be seen, the error decreases rapidly with the grid resolution, falling below $1 \%$ for $\Delta^{-1} \geq 2$. 
Altogether, this shows that our simulation results are not appreciably affected by finite resolution effects (up to an error of about $1 \%$ for $\Delta = \frac{1}{2}$).

We also have checked that our simulation results are not constrained by the positions of the inlet and outlet. To this end, we have performed additional simulations, putting inlet and outlet twice as far from the metric perturbations, observing that our simulation results do not change with the inlet and outlet positions (if they are sufficiently apart as in all the simulations presented in the paper).

\subsection{Soap film experiment}

The soap film in Fig. 1 in the Letter has been created by spanning a film, made of a commercial soap bubble solution, between two parallel vertical wires like in Ref. \cite{zhang2000flexible}. The flowing soap film is driven by gravity, and an out-of-plane curvature can be achieved by applying a bundled air flow normal to the film surface. Out-of-plane curvature can also be achieved by applying an electrostatic field (e.g. by a charged balloon) or by bending the confining wires.

\subsection{Permeability fitting functions}

Table \ref{tab:fitting-functions} contains the fitting functions corresponding to the permeability plots in Fig. 5 in the Letter.

\begin{table}[ht]
  \centering 
  \begin{tabular}{|ccc|}\hline
    $\a(n)$ &$=$& $(5.6 \pm 0.5) \times 10^{-2} n - (8.9 \pm 3.9) \times 10^{-5}$ \\
    $\b(n)$ &$=$& $(1.9 \pm 0.4) \times 10^{-2} n - (5.0 \pm 3.8) \times 10^{-5}$ \\
    $\a(a_0)$ &$=$& $(1.07 \pm 0.04) \times 10^{-3} a_0 - (5.2 \pm 0.3) \times 10^{-5}$ \\
    $\b(a_0)$ &$=$& $(8.9 \pm 2.6) \times 10^{-4} a_0 - (2.4 \pm 2.0) \times 10^{-5}$ \\
    $\a(\epsilon)$ &$=$& $(1.1 \pm 0.8) \times 10^{-4} \epsilon^2 - (4.5 \pm 0.7) \times 10^{-5}$ \\
    $\b(\epsilon)$ &$=$& $(5.7 \pm 1.3) \times 10^{-5} \epsilon^2 - (4.1 \pm 1.4) \times 10^{-5}$ \\
    $\a(K)$ &$=$& $(1.05 \pm 0.03) \times 10^{-1} K - (2.8 \pm 0.8) \times 10^{-5}$ \\
    $\b(K)$ &$=$& $(4.3 \pm 0.6) \times 10^{-2} K - (3.5 \pm 1.6) \times 10^{-5}$
    \\ \hline
  \end{tabular}
  \caption{Fitting functions for the reciprocal permeabilities $\a$ and $\b$.} 
  \label{tab:fitting-functions}  
\end{table}

\bibliography{citations}

\end{document}